\documentclass[12pt]{iopart}

\usepackage[pdftex]{graphicx}
\usepackage[nodvipsnames]{color}
 
\begin{document}

\title[The neutron cross section of low-temperature heteronuclear diatomic fluids]{The neutron cross section of low-temperature heteronuclear diatomic fluids}

\author{ Eleonora Guarini}

\address{Dipartimento di Fisica e Astronomia, Universit\`a degli Studi di Firenze, via G. Sansone 1, I-50019 Sesto Fiorentino, Italy}
\ead{guarini@fi.infn.it}
\vspace{10pt}
\begin{indented}
\item[]April 2021
\end{indented}

\begin{abstract}
The present work deals with the formal description of the response to neutrons of heteronuclear diatomic liquids, with special interest in the case of hydrogen deuteride as a possible candidate for the moderation process required in the production of cold neutrons. Preliminary evaluations of the model giving the neutron double differential cross section of a heteronuclear vibrating rotor are performed by using, as a first approximation, the ideal gas law for the centre-of-mass translational dynamics, which is expected to be appropriate at incident neutron energies above the thermal region. The unavailability of double differential cross section experimental data on liquid HD compels to test the model calculations only at an integral level, i.e. against the only measurement carried out on liquid HD for the determination of its neutron total cross section. The present findings indicate the evident need of more accurate measurements of the total cross section, as well as of future work devoted to double differential cross section determinations and appropriate simulations of the translational dynamics of this weakly quantum fluid.
\end{abstract}

%
%
%
%
%

\normalsize

\section{Introduction}

Liquid hydrogen and its isotopic forms are among the most used cryogenic fluids and, in the specific application to neutron techniques, are the most important low-temperature moderators used to realize cold neutron sources. In recent years, quite an effort has been devoted to deeply refine the ability to predict the neutron scattering properties of these diatomic fluids and build up reliable databases collecting their double differential cross section (DDCS) in as wide as possible kinematic ranges. As far as hydrogen (H$_2$) and deuterium (D$_2$) are concerned, a significant progress has been possible a few years ago when crucial experimental work has been dedicated to accurate measurements of the total cross section (TCS) of para-H$_2$ \cite{Grammer2015} and, at the same time, quantum calculation methods have been employed, in place of classical evaluations of the translational dynamics, in the DDCS algorithms for these low-mass molecules undergoing nonnegligible delocalisation effects \cite{Guarini2015, Guarini2016}. Very recently, the latter approach, which gives the unique opportunity to get rid of adjustable parameters in DDCS evaluations, has been successfully adopted in moderator design and nuclear data processing codes \cite{DamianHD2021} implemented at the European Spallation Source (ESS, Sweden). Therefore, concerning the homonuclear representatives of the hydrogen family, we can be rather satisfied with the capabilities acquired from a scientific point of view, now permitting well-grounded and safe applications to forthcoming and existing neutron sources. Further work is anyway in progress for a better assessment of the coherent scattering properties of D$_2$, mainly based on absolute scale determinations of its DDCS by inelastic neutron scattering experiments \cite{Guarini20XX}. 

Nonetheless, another member among the hydrogen liquids, namely hydrogen deuteride (HD), has not deserved in the years the same attention of its homonuclear partners, either from a scientific point of view, or in applications, although some of its midway properties between H$_2$ and D$_2$ might be conveniently exploited also in neutron moderation processes, as suggested in Ref.\ \cite{DamianHD2021}. For instance, it combines a still low absorption (i.e., an advantageous property typical of D$_2$) with a still interesting low-mass-related moderation efficiency (i.e., the key property of H$_2$), and all of this at the same (manageable) low temperatures of liquid H$_2$ and D$_2$ ($\sim$ 20 K). On the other hand, it is not clear how technical difficulties related to the long term instability of HD could be circumvented with sustainable costs. Nevertheless, when exploring the literature about experimental, theoretical and simulation work on liquid HD and, in particular, its neutron scattering properties, one is compelled to face a sort of unexpected desert. It is also quite surprising that the neutron response of HD, which anyway belongs to an important class of systems, escaped for decades a detailed treatment like those elegantly devised on homonuclear diatomic molecules by Young and Koppel (YK) \cite{YoungKoppel1964} and by Sears \cite{Sears1966} in the mid 1960s. Indeed, only much later the neutron cross section of HD has been briefly taken into consideration in the analysis of solid state data \cite{Colognesi2009}, and further years passed before a formal description of the heteronuclear diatomic case was tackled by the research group studying the behaviour of HD in the cages of clathrate hydrates \cite{Xu2013,Colognesi2014}. However, like the papers by Sears on the homonuclear case \cite{Sears1966}, also these last works deal with the scattering of cold and thermal neutrons from a low-temperature sample, therefore vibrations are not excited and the developed formalism is rightly limited to the specific case under consideration, where rotations alone are excited and only zero-point vibrational effects must be retained.    

In order to extend the applicability of the model also to the higher incident energies involved in applications, we therefore found it important to provide a more general treatment of heteronuclear diatomic fluids, including harmonic vibrations and overcoming the use of Debye-Waller factors \cite{Colognesi2009}, as formalised for the homonuclear case in Refs.\ \cite{YoungKoppel1964,Zoppi93} and like we did in Refs.\cite{Guarini2015, Guarini2016, Guarini2003, Crisp_report}. In summary, here we provide the corresponding formalism for the DDCS of a heteronuclear (harmonically) vibrating rotor.

To this aim, it is useful to recall under which hypotheses the adopted modeling of the DDCS holds \cite{YoungKoppel1964,Zoppi93,Guarini2003}: i) molecules are considered to be {\it free} vibrating rotors; ii) the translational centre of mass (CM) dynamics is assumed to be completely decoupled from the intramolecular motions; iii) rotations are treated as independent of vibrations and viceversa; iv) anharmonicity effects are neglected. The first hypothesis corresponds to consider an isotropic interaction among the molecules, with negligible orientational correlations. Moreover, for the cases of special interest like the one of low temperature liquids, an additional assumption is that all the molecules in the system lie initially (that is, before interaction with neutrons) in the ground vibrational state. Finally, only the case of unpolarised neutrons is considered here. 

The above general assumptions for the treatment of the neutron cross sections of diatomic low temperature fluids are further accompanied by the important simplifications introduced by the heteronuclear nature of the molecule, which allows to consider its nuclei as distinguishable particles. The absence of symmetry requirements for the total molecular wave function implies that there is no coupling between the total molecular spin and the rotational state of the molecule. Therefore, one can refer to the so-called ``uncorrelated spin'' case \cite{Guarini2003} where quantum-statistical averages involving nuclear spin variables can be carried out separately from those related to position variables. In what follows we will often resort to the contents, formulas, and references contained in Ref.\ \cite{Guarini2003}, which in the following will be referred to as paper I.

\section{Formalism}

\begin{figure}
\resizebox{0.7\textwidth}{!}
{\includegraphics[trim=-2cm 13cm 7cm 1cm]{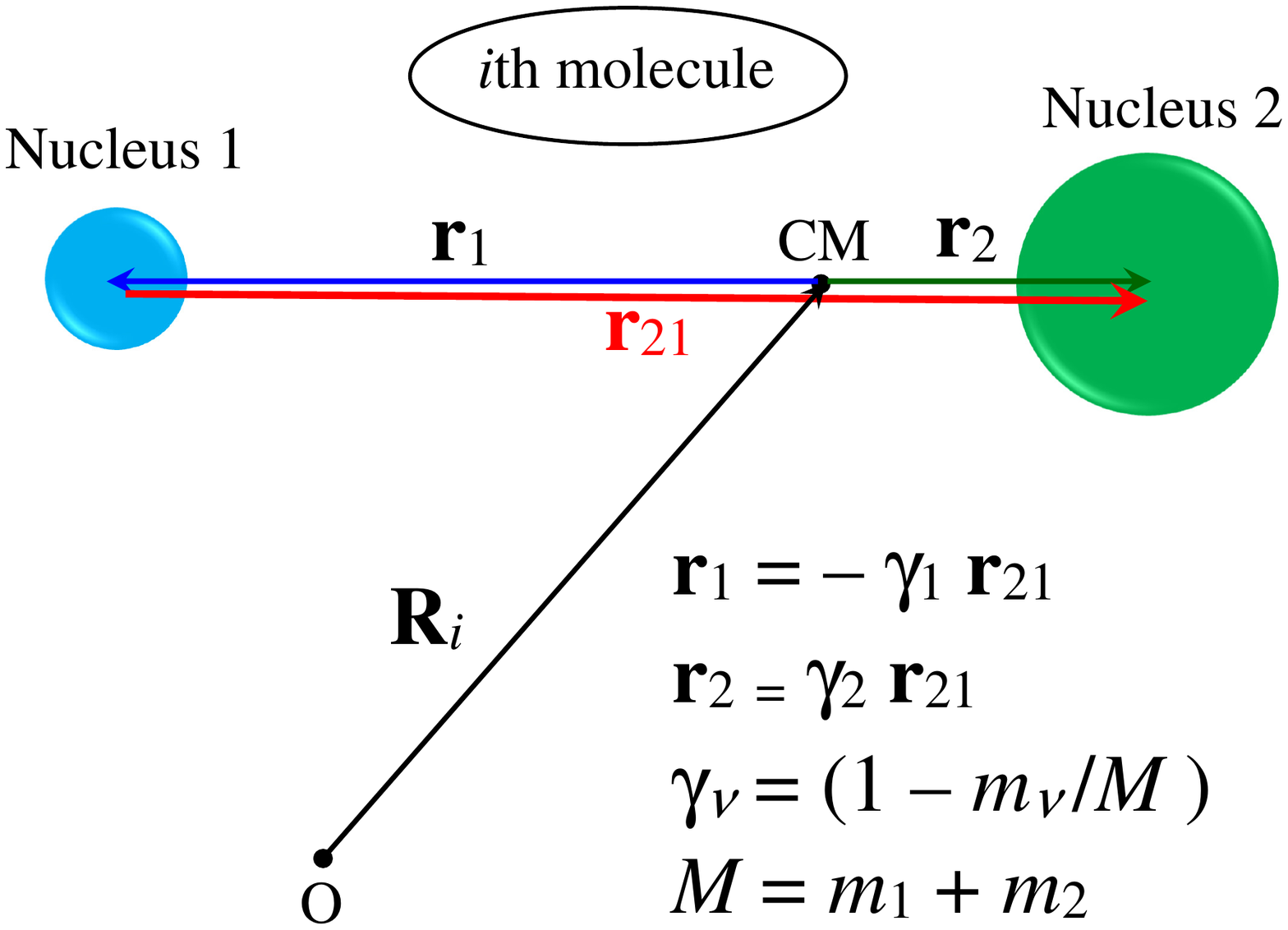}}
\caption{Geometry and position vectors defined for the case of a heteronuclear diatomic molecule. The positions of the nuclei with respect to the CM are expressed by the appropriate fractions $\gamma_1$ and $\gamma_2$ of the internuclear distance vector ${\bf r}_{21}$ (red arrow) joining the two nuclei of mass $m_1$ and $m_2$, respectively. $M$ is the total molecular mass. Note that $r_{21}$ is meant to represent the instantaneous value of the internuclear distance, which can further be written as $r_{21}=r_{\rm eq}+x$, with $x$ the bond stretching and $r_{\rm eq}$ the average equilibrium distance.}
\label{geom} 
\end{figure}

In the mentioned hypotheses and conditions, the basic starting formulas leading to the double differential cross section are those gathered in Eq.(7) of I which, by omitting the superscript {\it uncorr} there, we rewrite as:
\begin{eqnarray}
\label{d2sig_uncorr_generica}
\frac{d^2\sigma}{d\Omega d\omega} =\frac{k_1}{k_0}S_{\rm n}(Q,\omega)=\frac{k_1}{k_0} \frac{1}{2 \pi}\int{dt~ e^{-i \omega t}F_{\rm n}(Q,t)}=\\
\nonumber =\frac{k_1}{k_0} \frac{1}{2 \pi}\int{dt~ e^{-i \omega t}[u(Q) F_{\rm d}(Q,t)+v(Q,t) F_{\rm s}(Q,t)]},
\end{eqnarray}

\noindent where $k_0$ and $k_1$ are the incident and scattered neutron wavevectors, and $S_{\rm n}(Q,\omega)$ is the total dynamic structure factor provided by neutron scattering, i.e. the time Fourier transform of $F_{\rm n}(Q,t)$, which is the neutron weighted combination of the distinct and self intermediate scattering functions $F_{\rm d}(Q,t)$ and $F_{\rm s}(Q,t)$, respectively. For molecular fluids, the latter are to be identified with the CM functions defined by

\begin{eqnarray}
\label{CM_dist}
F_{\rm d}(Q,t)=\frac{1}{N} \sum_{i=1}^{N} \sum_{j=1 (j \ne i)}^{N} \Bigg \langle e^{-i {\bf Q} \cdot {\bf R}_i(0)} e^{i {\bf Q} \cdot {\bf R}_j(t)}\Bigg \rangle\\
\label{CM_self}
F_{\rm s}(Q,t)=\frac{1}{N} \sum_{i=1}^{N}\Bigg \langle e^{-i {\bf Q} \cdot {\bf R}_i(0)} e^{i {\bf Q} \cdot {\bf R}_i(t)}\Bigg \rangle
\end{eqnarray}

\noindent which provide the total CM intermediate scattering function $F(Q,t)=F_{\rm d}(Q,t)+F_{\rm s}(Q,t)$. In the above equations, $N$ is the total number of molecules, ${\bf R}_i(0)$ is the CM position of the $i$th molecule in the (arbitrarily chosen) time origin, and ${\bf R}_j(t)$ is the CM position of a different molecule at a subsequent time $t$. The angle brackets denote, as usual, a quantum canonical ensemble average. The isotropy of the fluid actually makes these functions depend only on the modulus $Q$ of the exchanged wavevector ${\bf Q}={\bf k}_0-{\bf k}_1$. Furthermore, in the last member of Eq.\ (\ref{d2sig_uncorr_generica}) the functions $u(Q)$ and $v(Q,t)$ play the role of inter and intra molecular form factors weighting, respectively, the distinct and self CM dynamics. In analogy with the monatomic case, purely coherent scattering characterizes $u(Q)$, that is, only the so-called coherent part of the neutron cross section of the scattering unit probes the interparticle translational dynamics. As a consequence, $u(Q)$ contains exclusively the coherent scattering lengths of the various nuclei present in the molecule and is independent of time. Differently, the intramolecular form factor is a function of time, and generally depends on both the coherent and incoherent nuclear scattering lengths. In particular, as reported in Eq.(7) of I, these functions can be written as

\begin{eqnarray}
\label{u_uncorr}   
 u(Q)=\Bigg | \sum_{\nu=1}^{n} b_{\rm coh,\nu} \sum_{u_0} p_{u_0} \langle u_0 |e^{i {\bf Q} \cdot {\bf r}_{\nu}} |u_0 \rangle  \Bigg |^2 \\
 \label{v_uncorr}
v(Q,t)=\sum_{\nu,\nu'=1}^{n} (b_{\rm coh,\nu}b_{\rm coh,\nu'}+b^2_{\rm inc,\nu} \delta_{\nu,\nu'}) \times \\
\nonumber \times \sum_{u_0,u_1}p_{u_0}e^{i \omega_{u_0 u_1}t}\langle u_0 |e^{-i {\bf Q} \cdot {\bf r}_{\nu}}|u_1\rangle \langle u_1 |e^{i {\bf Q} \cdot {\bf r}_{\nu'}}|u_0\rangle,
\end{eqnarray}

\noindent where $b_{\rm coh,\nu} $ and $b_{\rm inc,\nu}$ are the coherent and incoherent scattering lengths of the $\nu$th nucleus in the molecule, which, in the most general case, is assumed to be characterized by a total of $n$ (not necessarily different) nuclei. In the above equations, ${\bf r}_{\nu}$ is the vector defining the position, at $t=0$, of the $\nu$th nucleus with respect to the CM of the molecule. Figure \ref{geom} summarizes the various definitions in the simple case we are interested in, i.e. $n=2$ with nucleus 1 (e.g. H) different from nucleus 2 (e.g. D). 

The form factors are seen to involve the calculation of matrix elements of the kind $\langle u_1 |e^{i {\bf Q} \cdot {\bf r}_{\nu}}|u_0\rangle$, where $| u\rangle$ denotes a generic rotovibrational state of the molecule. Subscripts 0 and 1 are used to indicate the initial (before scattering) and final (after scattering) molecular state, respectively. This of course can be written explicitly by adopting the usual notation for the rotational and vibrational quantum numbers, i.e. $|u\rangle=|J M v\rangle=|J M\rangle|v\rangle$, the last equality descending from the mentioned hypothesis of negligible coupling between rotations and (harmonic) vibrations. Equations (\ref{u_uncorr}) and (\ref{v_uncorr}) include the statistical average over the initial state probabilities governed by the Boltzmann thermal distribution. Since the molecules are assumed to lie initially in the ground vibrational state ($v_0=0$), it is possible to write $p_{u_0}=p_{J_0} p_{M_0} p_{v_0}=p_{J_0} p_{M_0}$. Finally, the time dependence in Eq.\ (\ref{v_uncorr}) comes simply from the Heisenberg representation of ${\bf r}_{\nu'}(t)$ as $\exp(i H t /\hbar)~ {\bf r}_{\nu'} \exp(-i H t /\hbar)$, where $H=H_{\rm rot}+H_{\rm vib}$ is the total (rotational plus vibrational) Hamiltonian, $\hbar$ is the reduced Planck constant, and we defined the transition frequency as $\omega_{u_0 u_1}=(E_{u_1}-E_{u_0})/\hbar$. For a diatomic molecule, the latter can obviously be written also as $\omega_{u_0 u_1}=\omega_{J_0 J_1}+\omega_{v_0 v_1}=\omega_{J_0 J_1}+v_1 \omega_{\rm v}$, with $\omega_{\rm v}$ the frequency of the harmonic oscillator of mass $\mu=(m_1 m_2)/M$, i.e. corresponding to the reduced mass of the two-body system of total mass $M=m_1+m_2$, and whose energy levels are given by $E_{\rm vib}=\hbar \omega_{\rm v} ( v+\frac{1}{2})$. The rotational energy levels of a free rotor are as usual given by $E_{\rm rot}=J(J+1)[B-DJ(J+1)]$, with $B$ the rotational constant and $D$ accounting for centrifugal distortion, both constants being expressed in units of energy. 

Therefore, we can explicitly write:

\begin{eqnarray}
\label{u_expl}   
 u(Q)=\Bigg | \sum_{\nu=1}^{2} b_{\rm coh,\nu} \sum_{J_0 M_0} p_{J_0} p_{M_0} \langle J_0 M_0 | \langle 0 |e^{i {\bf Q} \cdot {\bf r}_{\nu}} |0 \rangle |J_0 M_0 \rangle  \Bigg |^2 \\
 \label{v_expl}  
 v(Q,t)=\sum_{J_0 M_0} p_{J_0} p_{M_0} \sum_{J_1 M_1 v_1} \sum_{\nu,\nu'=1}^{2} a_{\nu,\nu'} \times\\
 \nonumber \times e^{i \omega_{J_0 J_1}t} e^{i v_1 \omega_{\rm v}t}\langle J_0 M_0 | \langle 0 |e^{-i {\bf Q} \cdot {\bf r}_{\nu}}|v_1 \rangle |J_1 M_1 \rangle \langle J_1 M_1| \langle v_1 |e^{i {\bf Q} \cdot {\bf r}_{\nu'}}|0\rangle |J_0 M_0\rangle,
\end{eqnarray} 

\noindent where $a_{\nu,\nu'}=b_{\rm coh,\nu}b_{\rm coh,\nu'}+b^2_{\rm inc,\nu} \delta_{\nu,\nu'}$ \cite{Sears1966} results from the average of the scattering length over neutron and nuclear spin states which, in the uncorrelated case, can be performed separately from those involving position dependent operators. 

The next subsections are devoted to the calculation of Eqs.\ (\ref{u_expl}) and (\ref{v_expl}). Since the former is easily derived as a special case of the latter, we first address the case of $v(Q,t)$.

\subsection{\label{intra}The intramolecular form factor $v(Q,t)$}

By following Fig.\ \ref{geom}, and using the synthetic notation $\langle f | ... | i\rangle$ for $\langle J_1 M_1| \langle v_1 | ... |0\rangle |J_0 M_0\rangle$, Eq.\ (\ref{v_expl}) becomes

\begin{eqnarray}
\nonumber v(Q,t)=\sum_{J_0 M_0} p_{J_0} p_{M_0} \sum_{J_1 M_1 v_1}  e^{i \omega_{J_0 J_1}t} e^{i v_1 \omega_{\rm v}t} \times \\
\nonumber \times \Bigg ( a_{11} | \langle f | e^{i {\bf Q} \cdot {\bf r}_1} | i\rangle |^2 +a_{22} | \langle f | e^{i {\bf Q} \cdot {\bf r}_2} | i\rangle |^2+\\
\nonumber+ a_{12} \langle i | e^{- i {\bf Q} \cdot {\bf r}_1} |f\rangle \langle f | e^{i {\bf Q} \cdot {\bf r}_2} | i\rangle+a_{21} \langle i | e^{- i {\bf Q} \cdot {\bf r}_2} |f\rangle \langle f | e^{i {\bf Q} \cdot {\bf r}_1} | i\rangle\Bigg )=\\
\label{v_expl2}
= \sum_{J_0 M_0} p_{J_0} p_{M_0} \sum_{J_1 M_1 v_1}  e^{i \omega_{J_0 J_1}t} e^{i v_1 \omega_{\rm v}t} \times \\
\nonumber \times \Bigg [ a_{11} | \langle f | e^{-i {\bf Q} \cdot \gamma_1{\bf r}_{21}} | i\rangle |^2 +a_{22} | \langle f | e^{i {\bf Q} \cdot \gamma_2 {\bf r}_{21}} | i\rangle |^2+\\
\nonumber+ a_{12} \Bigg( \langle i | e^{i {\bf Q} \cdot \gamma_1{\bf r}_{21}} |f\rangle \langle f | e^{i {\bf Q} \cdot \gamma_2{\bf r}_{21}} | i\rangle+\langle i | e^{- i {\bf Q} \cdot \gamma_2 {\bf r}_{21}} |f\rangle \langle f | e^{-i {\bf Q} \cdot \gamma_1 {\bf r}_{21}} | i\rangle \Bigg) \Bigg ],
\end{eqnarray}

\noindent where in the last member we introduced the internuclear vector ${\bf r}_{21}$, duly weighted, through the factor $\gamma_{\nu}=1-m_{\nu}/M$, by the mass of the nucleus under consideration. Note that in the adopted notation both $\gamma_1$ and $\gamma_2$ are positive. Moreover, $a_{12}=a_{21}=b_{\rm coh,1} b_{\rm coh,2}$, while $a_{jj}=b^2_{{\rm coh},j} + b^2_{{\rm inc},j}$ when $j=1,2$. 

All terms in Eq.\ (\ref{v_expl2}) require the evaluation of the generic matrix element $B_{j \pm}=\langle J_1 M_1| \langle v_1 |e^{\pm i {\bf Q} \cdot \gamma_j {\bf r}_{21}}|0\rangle |J_0 M_0\rangle$. For instance, the first two terms are straightforwardly obtained from $B_{j+}^* B_{j+} = B_{j-}^* B_{j-}=|B_{j\pm}|^2$, while the cross term in the last line of the equation requires the calculation of e.g. $D=\langle J_0 M_0| \langle 0 |e^{i {\bf Q} \cdot \gamma_1 {\bf r}_{21}}|v_1\rangle |J_1 M_1\rangle \langle J_1 M_1| \langle v_1 |e^{i {\bf Q} \cdot \gamma_2 {\bf r}_{21}}|0\rangle |J_0 M_0\rangle=B_{1-}^* B_{2+}$, so that the quantity in round brackets actually reduces to $D+D^*=2 ~{\rm Re}D$. 

\subsubsection{Vibrations}

We thus turn to the direct calculation of the vibrational matrix element given by

\begin{eqnarray}
\label{vib}
\langle v_1 |e^{i {\bf Q} \cdot \gamma_j {\bf r}_{21}}|0\rangle=\\
\nonumber =\langle v_1 |e^{i Q \gamma_j r_{21} \eta}|0\rangle=\langle v_1 |e^{i Q \gamma_j (r_{\rm eq}+x) \eta}|0\rangle=e^{i \gamma_j \beta \eta}\langle v_1 |e^{i Q \gamma_j x \eta}|0\rangle,
\end{eqnarray}

\noindent where we explicitly wrote the scalar product posing $\eta=\cos \theta$, $\theta$ being the angle between {\bf Q} and {\bf r}$_{21}$. The internuclear distance was also expressed as the sum of the equilibrium one, $r_{\rm eq}$, and the bond stretching $x$. Finally, we defined $\beta=Q r_{\rm eq}$.

In order to evaluate the last member of Eq.\ (\ref{vib}), we direct the reader to the quantum mechanical treatment of a one-dimensional harmonic oscillator \cite{Messiah} of mass $\mu$, which provides $x=\sqrt{\frac{\hbar}{2 \mu \omega_{\rm v}}}(a^\dag+a)$ in terms of the Bose creation and annihilation operators obeying the commutation relation $[a,a^\dag]=1$. Using the properties $e^{\rm A}e^{\rm B}e^{[B,A]/2}=e^{\rm A+B}$ (which is valid in the present case) and $a|0\rangle=0$, the exponential of an operator, and considering the way $a^\dag$ operates on vibrational levels (leading e.g. to $\sqrt{v!}~|v\rangle=(a^\dag)^v |0\rangle$), it is possible to find that

\begin{eqnarray}
\label{vib2}
\langle v_1 |e^{i {\bf Q} \cdot \gamma_j {\bf r}_{21}}|0\rangle=\\
\nonumber  e^{i \gamma_j \beta \eta}\langle v_1 |e^{i Q \gamma_j x \eta}|0\rangle=e^{i \gamma_j \beta \eta} e^{-\frac{(\gamma_j \alpha \eta)^2}{2}} \frac{(i \gamma_j \alpha \eta)^{v_1}}{\sqrt{v_1!}}=\frac{(i \gamma_j \alpha)^{v_1}}{\sqrt{v_1!}}f_j(\eta)
\end{eqnarray}

\noindent where we defined $\alpha=Q \sqrt{\hbar M/(2 m_1 m_2 \omega_{\rm v})}$ and $f_j(\eta)=e^{-\frac{(\gamma_j \alpha \eta)^2}{2}} e^{i \gamma_j \beta \eta} ~\eta^{v_1}$.

\subsubsection{Rotations}

The next step consists in the calculation of the rotational matrix element 

\begin{equation}
\langle J_1 M_1 |\frac{(i \gamma_j \alpha)^{v_1}}{\sqrt{v_1!}}f_j(\eta) |J_0 M_0\rangle=\int d\Omega (-1)^{M_1}Y_{J_1, -M_1} \frac{(i \gamma_j \alpha)^{v_1}}{\sqrt{v_1!}}f_j(\eta) Y_{J_0, M_0}
\label{rot}
\end{equation}

\noindent where we introduced the spherical harmonics \cite{Messiah} omitting for brevity their argument $(\theta,\phi)$, and integration is performed over the solid angle with $d\Omega=d\phi d\theta \sin \theta$. Well known properties (see e.g. \cite{GG}) can be used to rewrite Eq.\ (\ref{rot}) in terms of appropriate Clebsch-Gordan coefficients and Legendre polynomials. In particular, 

\begin{eqnarray}
\label{rot2}
\int d\Omega (-1)^{M_1}Y_{J_1, -M_1}f_j(\eta) Y_{J_0, M_0}=\\
\nonumber =\int d\Omega (-1)^{M_1}f_j(\eta) \sum_{l,m} \sqrt{\frac{(2 J_1+1)(2 J_0+1)}{4 \pi (2 l+1)}} C(J_1 J_0 l;000) C(J_1 J_0 l;-M_1 M_0 m) Y_{l m}
\end{eqnarray}

\noindent with $|J_1-J_0| \le l \le J_1+J_0.$ 
\noindent Moreover, since 

\begin{equation}
\int_0^{2 \pi} d\phi ~Y_{lm}(\theta,\phi)=2 \pi \delta_{m0} \sqrt{\frac{2l+1}{4 \pi}}P_l(\cos \theta),
\end{equation}
 
\noindent we also derive

\begin{eqnarray}
\label{rot3}
\int d\Omega (-1)^{M_1}Y_{J_1, -M_1}f_j(\eta) Y_{J_0, M_0}=\\
\nonumber =\int_{-1}^1 d\eta \frac{(-1)^{M_1}}{2} \sqrt{(2 J_1+1)(2 J_0+1)} f_j(\eta) \sum_{l} C(J_1 J_0 l;000) C(J_1 J_0 l;-M_1 M_0 0) P_{l}(\eta).
\end{eqnarray}

\noindent Therefore

\begin{eqnarray}
\nonumber B_{j+}=\frac{(-1)^{M_1}}{2} \sqrt{(2 J_1+1)(2 J_0+1)} \sum_{l} C(J_1 J_0 l;000) C(J_1 J_0 l;-M_1 M_0 0) \times\\
\times \frac{(i \gamma_j \alpha)^{v_1}}{\sqrt{v_1!}} \int_{-1}^1 d\eta  f_j(\eta) P_{l}(\eta) =B_{j-}^*
\label{rot4}
\end{eqnarray}

\noindent We can finally calculate the {\it self-atomic} terms in Eq.\ (\ref{v_expl2}) according to

\begin{eqnarray}
\label{selfterm}
a_{jj} \sum_{J_0 M_0} p_{J_0} p_{M_0} \sum_{J_1 M_1 v_1}  e^{i \omega_{J_0 J_1}t} e^{i v_1 \omega_{\rm v}t} |B_{j+}|^2 = \\
\nonumber =a_{jj}  \sum_{J_0 J_1 v_1} p_{J_0} e^{i \omega_{J_0 J_1}t} e^{i v_1 \omega_{\rm v}t} \frac{(2J_1+1) (\gamma_j \alpha)^{2v_1}}{4 v_1!} \sum_l C^2(J_1 J_0 l;000) \Bigg| \int_{-1}^1 d\eta  f_j(\eta) P_{l}(\eta) \Bigg|^2,
\end{eqnarray}

\noindent where we exploited the property \cite{GG}

\begin{equation}
\sum_{M_0 M_1} p_{M_0}C(J_1 J_0 l;-M_1 M_0 0) C(J_1 J_0 l';-M_1 M_0 0)=\frac{\delta_{ll'}}{2J_0+1}.
\end{equation}

\noindent In analogy with I, we can here define the slightly more generalized integrals $A^{(j)}_{l v_1}$ as

\begin{equation}
 A^{(j)}_{l v_1}=\int_{-1}^1 d\eta ~e^{-\frac{(\gamma_j \alpha \eta)^2}{2}} e^{i \gamma_j \beta \eta} ~\eta^{v_1} P_{l}(\eta)=\int_{-1}^1 d\eta ~f_j(\eta) P_{l}(\eta),
 \label{al}
\end{equation}

\noindent which will be used in what follows to shorten the notation. Note that the first exponential in the integrand is an even function of $\eta$. So, if the complex exponential is split according to Euler formula, the former does not alter the well defined parity of the remaining product function.  

As concerns the {\it distinct-atomic} term in the last row of Eq.\ (\ref{v_expl2}), and considering that Eq.\ (\ref{rot4}) implies
\begin{eqnarray}
\nonumber D=B_{1+}B_{2+}=\frac{(-1)^{v_1}(\gamma_1\gamma_2)^{v_1}(\alpha)^{2v_1}}{4 v_1!} (2 J_1+1)(2 J_0+1) \times \\\times \sum_{ll'} C(J_1 J_0 l;000) C(J_1 J_0 l;-M_1 M_0 0) \int_{-1}^1 d\eta  f_1(\eta) P_{l}(\eta) \times\\
\nonumber \times C(J_1 J_0 l';000) C(J_1 J_0 l';-M_1 M_0 0) \int_{-1}^1 d\eta  f_2(\eta) P_{l'}(\eta),
\end{eqnarray}

\noindent one finds

\begin{eqnarray}
\nonumber a_{12} \sum_{J_0 M_0} p_{J_0} p_{M_0} \sum_{J_1 M_1 v_1}  e^{i \omega_{J_0 J_1}t} e^{i v_1 \omega_{\rm v}t} (D+D^*) = \\
\label{cross}
=a_{12}  \sum_{J_0 J_1 v_1} p_{J_0} e^{i \omega_{J_0 J_1}t} e^{i v_1 \omega_{\rm v}t} \frac{(2J_1+1) (-1)^{v_1}(\gamma_1 \gamma_2)^{v_1} \alpha^{2v_1}}{4 v_1!} \times\\
\nonumber \times \sum_l C^2(J_1 J_0 l;000) \Bigg[A^{(1)}_{lv_1}A^{(2)}_{lv_1}+A^{(1)*}_{lv_1}A^{(2)*}_{lv_1}\Bigg],
\end{eqnarray}

\noindent where, remembering Eq.\ (\ref{al}), the square bracket in the above equation can also be written as

\begin{eqnarray}
\nonumber \Bigg[...\Bigg]=2 \Bigg [\int_{-1}^1 d\eta...\cos(\gamma_1\beta\eta)\eta^{v_1}P_l(\eta)\int_{-1}^1 d\eta...\cos(\gamma_2\beta\eta)\eta^{v_1}P_l(\eta)] +\\
\label{quadra}
-\int_{-1}^1 d\eta...\sin(\gamma_1\beta\eta)\eta^{v_1}P_l(\eta)\int_{-1}^1 d\eta...\sin(\gamma_2\beta\eta)\eta^{v_1}P_l(\eta) \Bigg] =\\
=\nonumber 2 \Bigg [{\rm Re}A^{(1)}_{lv_1}{\rm Re}A^{(2)}_{lv_1}- {\rm Im}A^{(1)}_{lv_1}{\rm Im}A^{(2)}_{lv_1}\Bigg].
\end{eqnarray}

\subsubsection{\bf Final expression for $v(Q,t)$}
\label{finalv}
Combining Eqs.\ (\ref{v_expl2}), (\ref{selfterm}), (\ref{al}), (\ref{cross}), and (\ref{quadra}), the intramolecular form factor of a heteronuclear diatomic fluid turns out to be

\begin{eqnarray}
\nonumber v(Q,t)=\sum_{J_0 J_1 v_1} p_{J_0} e^{i \omega_{J_0 J_1}t} e^{i v_1 \omega_{\rm v}t} \frac{(2J_1+1) \alpha^{2v_1}}{4 v_1!} \sum_l C^2(J_1 J_0 l;000) \times \\
\label{vQt1}
\times \Bigg\{ (b^2_{{\rm coh},1} + b^2_{{\rm inc},1})\gamma_1^{2v_1}| A^{(1)}_{lv_1}|^2+(b^2_{{\rm coh},2} + b^2_{{\rm inc},2})\gamma_2^{2v_1}| A^{(2)}_{lv_1}|^2+\\
\nonumber+2~ b_{\rm coh,1} b_{\rm coh,2} (-1)^{v_1} (\gamma_1 \gamma_2)^{v_1} \Bigg[{\rm Re}A^{(1)}_{lv_1}{\rm Re}A^{(2)}_{lv_1}- {\rm Im}A^{(1)}_{lv_1}{\rm Im}A^{(2)}_{lv_1}\Bigg]\Bigg\}
\end{eqnarray}

\noindent In order to achieve a more general expression of $v(Q,t)$, which becomes particularly compact in the special case of a homonuclear diatomic fluid, it is useful to study the behaviour of the last row in Eq.\ (\ref{vQt1}) with varying the parity of $v_1$ and $l$, keeping in mind that the Legendre polynomials of order $l$ have the same parity of $l$. We thus distinguish the following cases: 

\begin{itemize}
\item $v_1 ~even$ - In this case $\eta^{v_1}$ is even, so:
\begin{equation}
\label{v1even}
\cases{{\rm Im}A^{(1)}_{lv_1}{\rm Im}A^{(2)}_{lv_1}=0&for $l~even$\\
{\rm Re}A^{(1)}_{lv_1}{\rm Re}A^{(2)}_{lv_1}=0&for $l~odd$\\}
\end{equation}
This implies that 
\begin{equation}
(-1)^{v_1} [...]=\cases{{\rm Re}A^{(1)}_{lv_1}{\rm Re}A^{(2)}_{lv_1}=\\ =\Bigg[{\rm Re}A^{(1)}_{lv_1}{\rm Re}A^{(2)}_{lv_1}+ {\rm Im}A^{(1)}_{lv_1}{\rm Im}A^{(2)}_{lv_1}\Bigg]&for $l~even$\\
-{\rm Im}A^{(1)}_{lv_1}{\rm Im}A^{(2)}_{lv_1}=\\=-\Bigg[{\rm Re}A^{(1)}_{lv_1}{\rm Re}A^{(2)}_{lv_1}+ {\rm Im}A^{(1)}_{lv_1}{\rm Im}A^{(2)}_{lv_1}\Bigg]&for $l~odd$\\}
\end{equation}
\item $v_1 ~odd$ - In this case $\eta^{v_1}$ is odd, so:
\begin{equation}
\label{v1odd}
\cases{{\rm Re}A^{(1)}_{lv_1}{\rm Re}A^{(2)}_{lv_1}=0&for $l~even$\\
{\rm Im}A^{(1)}_{lv_1}{\rm Im}A^{(2)}_{lv_1}=0&for $l~odd$\\}
\end{equation}
This implies that 
\begin{equation}
(-1)^{v_1} [...]=\cases{{\rm Im}A^{(1)}_{lv_1}{\rm Im}A^{(2)}_{lv_1}=\\ =\Bigg[{\rm Re}A^{(1)}_{lv_1}{\rm Re}A^{(2)}_{lv_1}+ {\rm Im}A^{(1)}_{lv_1}{\rm Im}A^{(2)}_{lv_1}\Bigg]&for $l~even$\\
-{\rm Re}A^{(1)}_{lv_1}{\rm Re}A^{(2)}_{lv_1}=\\=-\Bigg[{\rm Re}A^{(1)}_{lv_1}{\rm Re}A^{(2)}_{lv_1}+ {\rm Im}A^{(1)}_{lv_1}{\rm Im}A^{(2)}_{lv_1}\Bigg]&for $l~odd$\\}
\end{equation}
\end{itemize}

\noindent The above analysis consequently shows that the same expressions hold, either for $l$ even or for $l$ odd, regardless of the parity of $v_1$. Therefore, one can equivalently write

\begin{equation}
(-1)^{v_1} [...]=(-1)^l \Bigg[{\rm Re}A^{(1)}_{lv_1}{\rm Re}A^{(2)}_{lv_1}+ {\rm Im}A^{(1)}_{lv_1}{\rm Im}A^{(2)}_{lv_1}\Bigg]~{\rm for} ~ any ~v_1
\end{equation}

In conclusion, Eq.\ (\ref{vQt1}) can also be written in a way that disentangles the roles played by $v_1$ and $l$, and is governed exclusively by the parity of $l$, i.e.

\begin{eqnarray}
\nonumber v(Q,t)=\sum_{J_0 J_1 v_1} p_{J_0} e^{i \omega_{J_0 J_1}t} e^{i v_1 \omega_{\rm v}t} \frac{(2J_1+1) \alpha^{2v_1}}{4 v_1!} \sum_l C^2(J_1 J_0 l;000) \times \\
\label{vQt2}
\times \Bigg\{ (b^2_{{\rm coh},1} + b^2_{{\rm inc},1})\gamma_1^{2v_1}| A^{(1)}_{lv_1}|^2+(b^2_{{\rm coh},2} + b^2_{{\rm inc},2})\gamma_2^{2v_1}| A^{(2)}_{lv_1}|^2+\\
\nonumber +2~ b_{\rm coh,1} b_{\rm coh,2} (-1)^{l} (\gamma_1 \gamma_2)^{v_1} \Bigg[{\rm Re}A^{(1)}_{lv_1}{\rm Re}A^{(2)}_{lv_1}+ {\rm Im}A^{(1)}_{lv_1}{\rm Im}A^{(2)}_{lv_1}\Bigg]\Bigg\},
\end{eqnarray}

\noindent where we recall that, by definition, $\gamma_1 \gamma_2>0$.

As mentioned, the advantage of such a formulation for $v(Q,t)$ becomes evident in the homonuclear case, for which $m_1=m_2=m$, $\gamma_1=\gamma_2=\gamma=1/2$, $A^{(1)}_{lv_1}=A^{(2)}_{lv_1}=A_{lv_1}$, $b_{\rm coh,1}=b_{\rm coh,2}=b_{\rm coh}$, $b_{\rm inc,1}=b_{\rm inc,2}=b_{\rm inc}$, and Eq.\ (\ref{vQt2}) can be cast in the elegant form

\begin{eqnarray}
\nonumber v(Q,t)_{\rm homo}=\sum_{J_0 J_1 v_1} p_{J_0} e^{i \omega_{J_0 J_1}t} e^{i v_1 \omega_{\rm v}t} \frac{(2J_1+1) (\gamma\alpha)^{2v_1}}{4 v_1!} \sum_l C^2(J_1 J_0 l;000) \times \\
\times 2 \Bigg[ (b^2_{{\rm coh}} + b^2_{{\rm inc}})+ (-1)^{l} b^2_{\rm coh}\Bigg] |A_{lv_1}|^2,
\label{vQtomo}
\end{eqnarray} 

\noindent which is {\it effectively} identical to Eq.\ (31) given in I for the uncorrelated spin case. We specified {\it effectively} because, for the ease of notation here, {\it formal} differences appear between the formulas in I and the present ones. In fact, the quantities $\alpha$
(when $m_1=m_2=m$) and $\beta$ used here are not the same as in I. In particular, by renaming as $\alpha_{\rm I}$ and $\beta_{\rm I}$ those defined in paper I, we have

\begin{eqnarray}
\nonumber \alpha_{m_1=m_2=m}=Q\sqrt{\frac{\hbar}{m \omega_{\rm v}}}=2 \alpha_{\rm I}\\
\nonumber \beta=Q r_{\rm eq}=2 \beta_{\rm I}
\end{eqnarray} 

\noindent Nonetheless, Eqs.\ (\ref{al}) and (\ref{vQtomo}) only depend on the products $\gamma \alpha$ and $\gamma \beta$, that is, exactly on $\alpha_{\rm I}$ and $\beta_{\rm I}$ (since $\gamma=1/2$). 

\subsection{\label{inter}The intermolecular form factor $u(Q)$} 

In this case, the starting point is Eq.\ (\ref{u_uncorr}). Again, we represent the initial rotovibrational state of the molecule $|J_0 M_0\rangle|0\rangle$ with the synthetic notation $|i\rangle$, and explicitly write the square modulus in the equation, which becomes

\begin{eqnarray}
 \nonumber u(Q)=\sum_{\nu,\nu'=1}^{2} b_{\rm coh,\nu} b_{\rm coh,\nu'} \langle e^{-i {\bf Q} \cdot {\bf r}_{\nu}} \rangle \langle e^{i {\bf Q} \cdot {\bf r}_{\nu'}} \rangle =\\
 \nonumber  = b^2_{\rm coh,1} |\langle e^{i {\bf Q} \cdot {\bf r}_{1}} \rangle|^2+b^2_{\rm coh,2} |\langle e^{i {\bf Q} \cdot {\bf r}_{2}} \rangle|^2\\
  \label{u_expl2} 
b_{\rm coh,1} b_{\rm coh,2} \langle e^{-i {\bf Q} \cdot {\bf r}_{1}} \rangle \langle e^{i {\bf Q} \cdot {\bf r}_{2}} \rangle+ b_{\rm coh,2} b_{\rm coh,1} \langle e^{-i {\bf Q} \cdot {\bf r}_{2}} \rangle \langle e^{i {\bf Q} \cdot {\bf r}_{1}} \rangle=\\
\nonumber b^2_{\rm coh,1} |\langle e^{-i {\bf Q} \cdot {\gamma_1 \bf r}_{21}} \rangle|^2+b^2_{\rm coh,2} |\langle e^{i {\bf Q} \cdot {\gamma_2\bf r}_{21}} \rangle|^2\\
\nonumber  b_{\rm coh,1} b_{\rm coh,2} \langle e^{i {\bf Q} \cdot \gamma_1{\bf r}_{21}} \rangle \langle e^{i {\bf Q} \cdot \gamma_2{\bf r}_{21}} \rangle+b_{\rm coh,2} b_{\rm coh,1} \langle e^{-i {\bf Q} \cdot \gamma_2 {\bf r}_{21}} \rangle \langle e^{-i {\bf Q} \cdot\gamma_1 {\bf r}_{21}} \rangle
\end{eqnarray} 
 
\noindent where $\langle ... \rangle$ was used to indicate $\sum_{|i\rangle} p_{|i\rangle} \langle i |... |i \rangle$, and as done before we introduced the internuclear vector ${\bf r}_{21}$ in the last member of the equation. The calculation therefore consists in the evaluation of $\sum_{J_0 M_0}p_{J_0}p_{M_0}\langle J_0 M_0 |\langle0| e^{i {\bf Q} \cdot \gamma_j{\bf r}_{21}}|0\rangle|J_0 M_0\rangle$, which is easily carried out by considering the result reported in Eq.\ (\ref{rot4}) for $B_{j+}$ and by applying it to the case $v_1=0$, $|J_1 M_1\rangle=|J_0 M_0\rangle$. Doing so, we obtain

\begin{eqnarray}
\label{rot5}
\langle e^{i {\bf Q} \cdot \gamma_j{\bf r}_{21}}\rangle=\sum_{J_0 M_0}p_{J_0}p_{M0} \frac{(-1)^{M_0}}{2}(2 J_0+1) \times\\
\nonumber \times \sum_{l} C(J_0 J_0 l;000) C(J_0 J_0 l;-M_0 M_0 0) A^{(j)}_{l0}.
\end{eqnarray}

\noindent To further simplify Eq.\ (\ref{rot5}), we make use of the following relations \cite{GG}:

\begin{eqnarray}
\nonumber \sum_{M_0}p_{M0} C(J_0 J_0 l;-M_0 M_0 0)=(-1)^{J_0}\frac{\delta_{l0}}{\sqrt{2J_0+1}}\\
\nonumber C(J_0 J_0 0;0 0 0)=\frac{(-1)^{J_0}}{\sqrt{2J_0+1}}
\end{eqnarray}

\noindent which together provide

\begin{equation}
\langle e^{i {\bf Q} \cdot \gamma_j{\bf r}_{21}}\rangle=\sum_{J_0}p_{J_0}\frac{A^{(j)}_{00}}{2}=\frac{A^{(j)}_{00}}{2}.
\end{equation}

\noindent By inserting this last result in Eq.\ (\ref{u_expl2}), we get

\begin{eqnarray}
\label{u_expl3}   
u(Q)= \frac{b^2_{\rm coh,1} }{4} |A^{(1)}_{00}|^2+\frac{b^2_{\rm coh,2} }{4} |A^{(2)}_{00}|^2+\\
 \nonumber +\frac{b_{\rm coh,1} b_{\rm coh,2} }{4}\Bigg[A^{(1)}_{00}A^{(2)}_{00}+A^{(1)*}_{00}A^{(2)*}_{00} \Bigg],
\end{eqnarray} 

\noindent which, by means of Eq.\ (\ref{quadra}), leads to

\begin{eqnarray}
\label{u_expl4}   
u(Q)= \frac{b^2_{\rm coh,1} }{4} |A^{(1)}_{00}|^2+\frac{b^2_{\rm coh,2} }{4} |A^{(2)}_{00}|^2+\\
\nonumber +\frac{b_{\rm coh,1} b_{\rm coh,2} }{2}\Bigg[{\rm Re}A^{(1)}_{00}{\rm Re}A^{(2)}_{00}-{\rm Im}A^{(1)}_{00}{\rm Im}A^{(2)}_{00} \Bigg].
\end{eqnarray} 

\noindent The present case corresponds to the one analysed in subsection \ref{finalv} when $v_1$ and $l$ are both even, therefore ${\rm Im}A^{(1)}_{00}{\rm Im}A^{(2)}_{00}=0$. This again means that one can equivalently write 

\begin{eqnarray}
\label{finalu}   
u(Q)= \frac{b^2_{\rm coh,1} }{4} |A^{(1)}_{00}|^2+\frac{b^2_{\rm coh,2} }{4} |A^{(2)}_{00}|^2+\\
 \nonumber +\frac{b_{\rm coh,1} b_{\rm coh,2} }{2}\Bigg[{\rm Re}A^{(1)}_{00}{\rm Re}A^{(2)}_{00}+{\rm Im}A^{(1)}_{00}{\rm Im}A^{(2)}_{00} \Bigg],
\end{eqnarray} 

\noindent which, in the homonuclear case, adds up to

\begin{equation}
\label{uhomo}   
u(Q)_{\rm homo}=b^2_{\rm coh}|A_{00}|^2,
\end{equation} 

\noindent in agreement with Eq.\ (27) of I.

\section{\label{DDCS} The neutron DDCS of a diatomic fluid at low temperature}

The neutron DDCS is obtained by inserting Eqs.\ (\ref{vQt2}) and (\ref{finalu}) in Eq.\ (\ref{d2sig_uncorr_generica}). Before discussing reasonable modelings of the CM translational dynamics, we wish to point out a few basic facts that are significant for the comparison of calculations with, possibly available, neutron experimental spectra. 

The first regards what is actually accessed by experiments. It is well known that conventional neutron spectroscopy provides the Fourier transforms of space and time correlation functions. In the previous section, we introduced the time autocorrelation of the spatial Fourier transform of the microscopic density, i.e. the total intermediate scattering function $F(Q,t)$, separated into its {\it distinct} and {\it self} parts (see Eqs.\ (\ref{CM_dist}) and (\ref{CM_self})). The latter were used in particular to the describe the neutron version of the molecular scattering function as $F_{\rm n}(Q,t)=u(Q)F_{\rm d}(Q,t)+v(Q,t)F_{\rm s}(Q,t)$. Such a separation of the total $F_{\rm n}(Q,t)$, inherited from the formalism used to describe neutron scattering from monatomic fluids, i.e. $b^2_{\rm coh}F_{\rm d}(Q,t)+(b^2_{\rm coh}+b^2_{\rm inc})F_{\rm s}(Q,t)$, has the merit to highlight the relationship between coherent scattering and the distinct dynamics. This holds true also for molecular liquids, since we saw that $u(Q)$ only contains the coherent scattering lengths of the constituent atoms. However, despite its conceptual significance, the mentioned separation has no feedback from reality, since neutrons provide a different combination and can only be the probe of ``true'' correlation functions, such as $F(Q,t)$ and $F_{\rm s}(Q,t)$, differently from $F_{\rm d}(Q,t)$. This means that the self and distinct contributions to the total dynamics cannot be disentangled in a neutron measurement on a totally coherent sample. Conversely, it is incoherent scattering that provides an, as remarkable as exclusive, pathway to the self dynamics. In other words, the output of a neutron experiment on a monatomic sample with nonzero coherent and incoherent scattering lengths is actually $b^2_{\rm coh}F(Q,t)+b^2_{\rm inc}F_{\rm s}(Q,t)$, the molecular version of which is

\begin{equation}
\label{Fmol_neutrons}   
F_{\rm n}(Q,t)=u(Q)F(Q,t)+[v(Q,t)-u(Q)]F_{\rm s}(Q,t).
\end{equation}  

The second thing worth recalling concerns the spectral properties and the general features of the resulting DDCS. By completing the switching to Fourier $(Q,\omega)$ space, Eq.\ (\ref{Fmol_neutrons}) becomes 

\begin{eqnarray}
\nonumber S_{\rm n}(Q,\omega)=\frac{1}{2 \pi}\Bigg\{u(Q)\int_{-\infty}^{+\infty} dt~ e^{-i\omega t}F(Q,t)+\\
+\int_{-\infty}^{+\infty} dt~e^{-i\omega t} [v(Q,t)-u(Q)]F_{\rm s}(Q,t)\Bigg \}=\\
\label{Smol_neutrons}   
=u(Q)S(Q,\omega)+\Bigg[\frac{1}{2 \pi}\int_{-\infty}^{+\infty} dt~e^{-i\omega t} v(Q,t) F_{\rm s}(Q,t)\Bigg]-u(Q) S_{\rm s}(Q,\omega).
\end{eqnarray}  
 \noindent The first term is therefore related to the CM total dynamic structure factor we are interested in when studying the translational collective dynamics of the system. Concerning the self properties, it is seen instead that the dependence of $v(Q,t)$ on time prevents one from expressing the second term as directly proportional to the self dynamic structure factor $S_{\rm s}(Q,\omega)$. However, Eq.\ (\ref{vQt2}) shows that time enters $v(Q,t)$ only in exponential form, therefore it is possible to write
 
 \begin{eqnarray}
\nonumber \int_{-\infty}^{+\infty} dt~e^{-i\omega t} v(Q,t) F_{\rm s}(Q,t)=\\
\nonumber =\sum_{J_0 J_1 v_1} \int_{-\infty}^{+\infty} dt~ e^{-i (\omega-\omega_{J_0 J_1}-v_1 \omega_{\rm v})t} \mathcal{F}(Q;J_0 J_1 v_1) F_{\rm s}(Q,t) 
\end{eqnarray}  

\noindent where we put 
\begin{eqnarray}
\nonumber \mathcal{F}(Q;J_0 J_1 v_1) =p_{J_0} \frac{(2J_1+1) \alpha^{2v_1}}{4 v_1!} \sum_l C^2(J_1 J_0 l;000) \times \\
\nonumber \times \Bigg\{ (b^2_{{\rm coh},1} + b^2_{{\rm inc},1})\gamma_1^{2v_1}| A^{(1)}_{lv_1}|^2+(b^2_{{\rm coh},2} + b^2_{{\rm inc},2})\gamma_2^{2v_1}| A^{(2)}_{lv_1}|^2+\\
\nonumber +2~ b_{\rm coh,1} b_{\rm coh,2} (-1)^{l} (\gamma_1 \gamma_2)^{v_1} \Bigg[{\rm Re}A^{(1)}_{lv_1}{\rm Re}A^{(2)}_{lv_1}+ {\rm Im}A^{(1)}_{lv_1}{\rm Im}A^{(2)}_{lv_1}\Bigg]\Bigg\}.
\end{eqnarray}

\noindent Consequently, the DDCS finally reads

\begin{eqnarray}
\label{d2sig_finale}
\frac{d^2\sigma}{d\Omega d\omega} =\frac{k_1}{k_0}S_{\rm n}(Q,\omega)=\frac{k_1}{k_0} \Bigg [ u(Q) S(Q,\omega)+\\
\nonumber +\sum_{J_0 J_1 v_1}  \mathcal{F}(Q;J_0 J_1 v_1)S_{\rm s}(Q,\omega-\omega_{J_0 J_1}-v_1 \omega_{\rm v})-u(Q)S_{\rm s}(Q,\omega) \Bigg ].
\end{eqnarray}

\noindent Equation (\ref{d2sig_finale}) shows that the single-molecule contribution to the DDCS corresponds to a comb of lines centred at the frequencies of the possible rotovibrational transitions. These spectral components are therefore either central or shifted replicas of the lineshape describing the CM $S_{\rm s}(Q,\omega)$, with amplitudes ruled by the involved quantum numbers, the initial state probabilities    and the nuclear scattering lengths. 

Calculations of the DDCS of course require a modeling of both $S(Q,\omega)$ and $S_{\rm s}(Q,\omega)$ we are going to discuss in the next subsection. Here, it is worth recalling that comparison with experiment is only possible if the model lineshapes obey the detailed balance principle. Therefore, if classical, i.e. symmetric, models are used for the dynamic structure factors, these must be duly asymmetrised via multiplication by the factor $[n(\omega)+1] =\frac{\hbar \omega}{k_{\rm B}T} /[1-\exp(-\frac{\hbar \omega}{k_{\rm B}T})]$ prior to their inclusion in Eq.\ (\ref{d2sig_finale}), where $n(\omega)$ is the Bose factor \cite{Ashcroft} and $k_{\rm B}$ indicates the Boltzmann constant. Moreover, the finite energy resolution of spectroscopic data needs to be taken into account by performing comparisons with calculations only after these have been properly broadened by the experimental resolution function.  

\subsection{\label{models}Models of the translational dynamics} 

The present work focuses on diatomic systems which are still in the dense fluid phase at temperatures where other systems already reach solidification. Therefore, it mainly deals with a few but extremely important light molecular fluids: hydrogen and its isotopes. It is well known that the low mass (between 2 and 6 a.m.u.) and relatively low temperatures (e.g. around 20 {\rm K}) of molecular hydrogen and its isotopes in the liquid phase make the de Broglie thermal wavelength \cite{HansenMcDonald} $\Lambda=h/\sqrt{2 \pi M k_{\rm B}T}$ reach values of the order of the molecular size, while remaining inferior to the average intermolecular distance \cite{BellissimaJCP2019}. Therefore, quantum delocalisation of individual particles affects the static and dynamic CM properties of these systems with respect to classical behaviour, while indistinguishability can still be assumed to play a negligible role, thus justifying the use of Boltzmann statistics. 

These overall assumptions are commonly summarised by saying that hydrogens are {\it moderate} quantum fluids, if compared to the paradigmatic case of helium. From a practical point of view, such a mild quantum nature has been the rationale behind the undiscouraged development of simulation algorithms still based on the possibility to define trajectories in phase space, but aimed at capturing the nonclassical effects of particle delocalisation, at least on the simplest time correlation functions relevant to fluid dynamics. In this respect, several positive results were gathered in the last decades about the effectiveness of Centroid Molecular Dynamics (CMD) \cite{Cao1994,Jang1999,Voth2004} and Ring Polymer Molecular Dynamics (RPMD) \cite{Craig2004,Miller2005,Habershon2013} simulation methods for the prediction of the CM velocity autocorrelation function (VAF) of the hydrogen homonuclear liquids. Among these confirmations, it is of special relevance in the present context the good performance of RPMD or CMD VAF calculations in estimates of the total neutron cross section of both H${_2}$ \cite{Guarini2015} and D${_2}$ \cite{Guarini2016}. In particular, the above algorithms were used to get quantum compliant evaluations of the VAF which, combined with the Gaussian Approximation \cite{GA1,RahmanSingwi1962}, are able to provide, at present, the most reliable determination of the CM $S_{\rm s}(Q,\omega)$ of the mentioned homonuclear liquids. 

Such a satisfactory situation is however not general, as soon as one considers the heteronuclear representatives of molecular hydrogen. Indeed, no simulation study has been devoted to evaluations of the VAF or of other correlation functions of HD. At the same time, no experiments were performed to explore the dynamic structure of this liquid at the nanometer and picosecond scales. Only neutron TCS data, collected in the 1970s by Seiffert \cite{Seiffert_rep,Seiffert}, are available for liquid HD and for indirect tests of the neutron scattering law summarized in Eq.\ (\ref{d2sig_finale}).

The present absence of simulation and experimental work on liquid HD does not diminish the importance of performing first checks about our ability to calculate its DDCS and, through the double integration

\begin{equation}
\label{sigma}   
\sigma=\int_{\Omega} d\Omega \int_{-\infty}^{\omega_0}  ~d\omega \frac{d^2\sigma}{d\Omega d\omega},
\end{equation}  

\noindent its TCS $\sigma$ with varying incident neutron energy, since this effort would anyway help assessing the critical issues and envisaging possible improvements for applications to neutron moderation as those announced in Ref.\ \cite{DamianHD2021}. In such an attempt, a fundamental step consists in evaluating the performance of the simplest analytical model available for the translational dynamics: the ideal gas (IG) law. This model completely neglects interaction and provides $S_{\rm s}(Q,\omega)$ in the form

\begin{equation}
\label{IG}   
S_{\rm s}(Q,\omega)=\frac{1}{Q}\sqrt{\frac{M}{2 \pi k_{\rm B} T}}\exp \Bigg[-\frac{M}{2  k_{\rm B} T Q^2}\Bigg( \omega-\frac{\hbar Q^2}{2 M}\Bigg)^2\Bigg],
\end{equation}  

\noindent that is, the well known Gaussian lineshape with standard deviation $\sigma=\sqrt{\frac{k_{\rm B} T Q^2}{M}}$ and centred at the recoil frequency $\omega_{\rm r}=\hbar Q^2 /(2 M)$. It is this last property of the IG profile that makes it asymmetric and compliant with the detailed balance condition. Assuming IG behaviour for the CM dynamics corresponds to the DDCS modeling originally devised by Young and Koppel \cite{YoungKoppel1964}. 

The YK recipe for the neutron DDCS of H$_2$ and D$_2$ can of course be exploited also in conjunction with apparently less crude descriptions of $S_{\rm s}(Q,\omega)$, like for instance the Egelstaff and Schofield model \cite{Guarini2003,ES,Copley} aimed at interpolating between the low $Q$ (hydrodynamic) and high $Q$ (kinetic) regimes of the single-particle dynamics, i.e. between simple diffusion and free particle behaviour. However, this cleverly conceived model, while being effective for classical fluids (see e.g. the case of methane discussed in Ref.\ \cite{Guarini2003}), turns out to be somewhat lacking for mild quantum liquids, notwithstanding the due modifications \cite{EgelstaffSoper1980,Zetterstrom1996} applied to ensure its fulfillment of the first frequency moment sum rule $M^{(1)}(Q)=\int{d\omega ~\omega~ S_{\rm s}(Q,\omega)}=\omega_{\rm r}$. Indeed, another fundamental spectral property, i.e. the one regarding the second frequency moment, was shown in Refs.\ \cite{Guarini2015,Crisp_report} to be rather heavily missed by the modified Egelstaff and Schofield schematisation, even more than in the IG case. In addition, considering that TCS calculations based on the IG lineshape tend to compare better and better with experimental data as the neutron incident energy is increased above some tens of meV \cite{Crisp_report}, there is actually no need to resort to other models if interested in certain energy ranges. More importantly, it is seen in Fig.\ 10 of Ref.\ \cite{Crisp_report} that IG behaviour starts to be quite accurate for normal H$_2$ already at 10 meV incident energy. Differently, D$_2$ requires a more realistic modeling of the CM dynamics, pushing to incident energies above 50 meV the range of reliability of TCS estimates using uniquely the IG limit of $S(q,\omega)$ (see Fig.\ 14 of Ref.\ \cite{Crisp_report}). Likely, HD will show an intermediate  behaviour between H$_2$ and D$_2$ which we are going to inquire by means of IG based calculations as a function of the incident neutron energy $E_0$.

\section{\label{HD}The case of hydrogen deuteride (HD)} 

\begin{figure}
\resizebox{0.9\textwidth}{!}
{\includegraphics[trim=0cm 0cm 2cm 0cm]{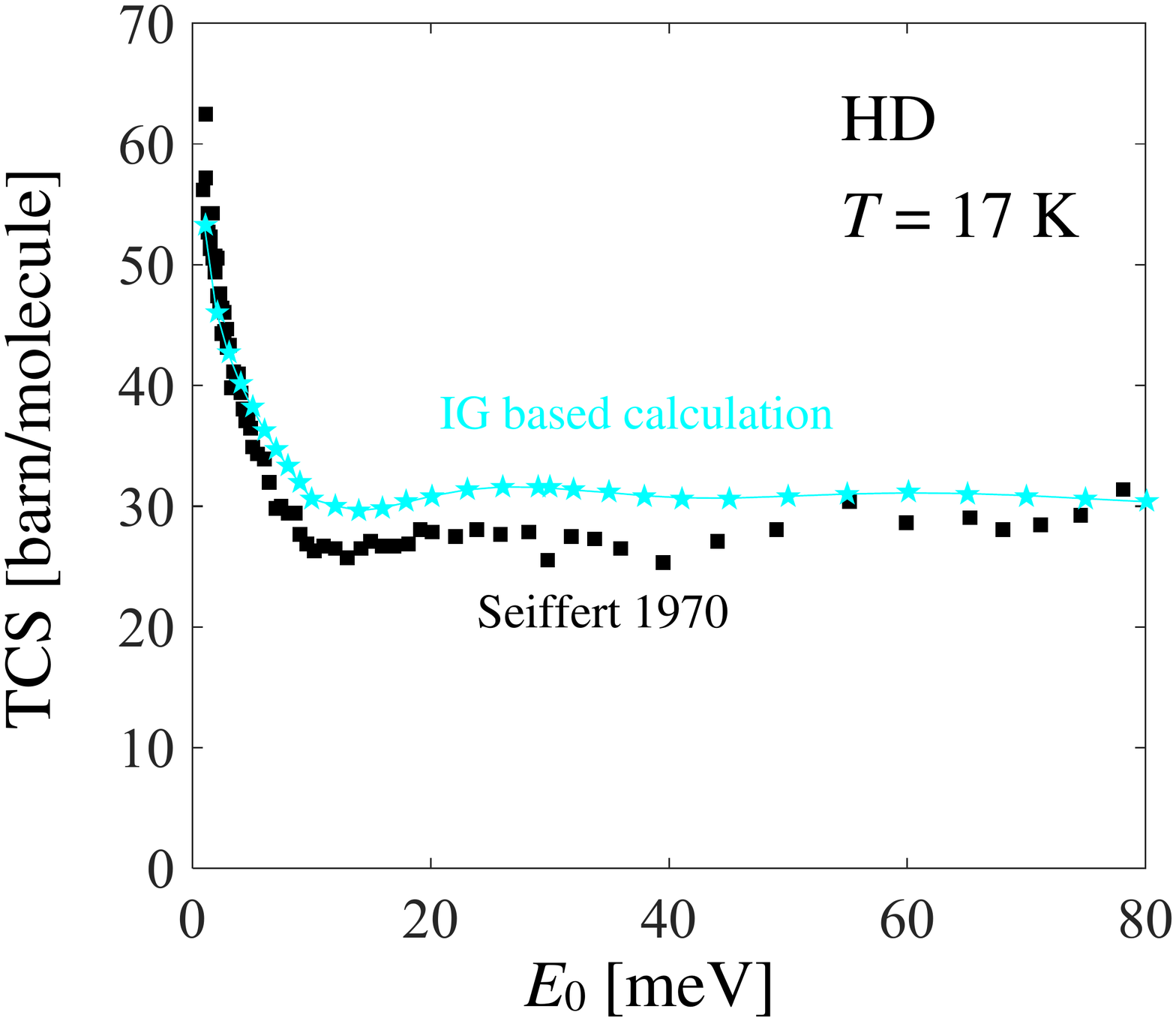}}
\caption{Dependence on the incident energy $E_0$ of the total neutron cross section of HD at 17 K as obtained by Seiffert \cite{Seiffert_rep} (black full circles) and by double integration of Eq.\ (\ref{d2sig_finale}) following Eq.\ (\ref{sigma}) (cyan stars with thin line), using the ideal gas law of Eq.\ (\ref{IG}) for the CM translational dynamics.}
\label{TCS} 
\end{figure}

Specific calculations for HD were performed by using the molecular parameters and neutron scattering lengths listed in Tab.\ \ref{Tab1}. The DDCS was calculated at $T=17$ K as a function of scattering angle and exchanged energy at various values of the incident energy $E_0$ ranging between 1 and 80 meV. At such energies only rotations are excited and the main contribution to the spectra comes from the elastic $J_0=0 \to J_1=0$ line and the $J_0=0 \to J_1=1$ Stokes transition (when excited) centred at about 11 meV. Double integration of the DDCS over solid angle and exchanged energy, according to Eq.\ (\ref{sigma}), provided the total scattering cross section shown in Fig.\ \ref{TCS}, where also Seiffert's data \cite{Seiffert_rep} are reported for comparison. 

\begin{figure}
\resizebox{0.9\textwidth}{!}
{\includegraphics[trim=0cm 0cm 2cm 0cm]{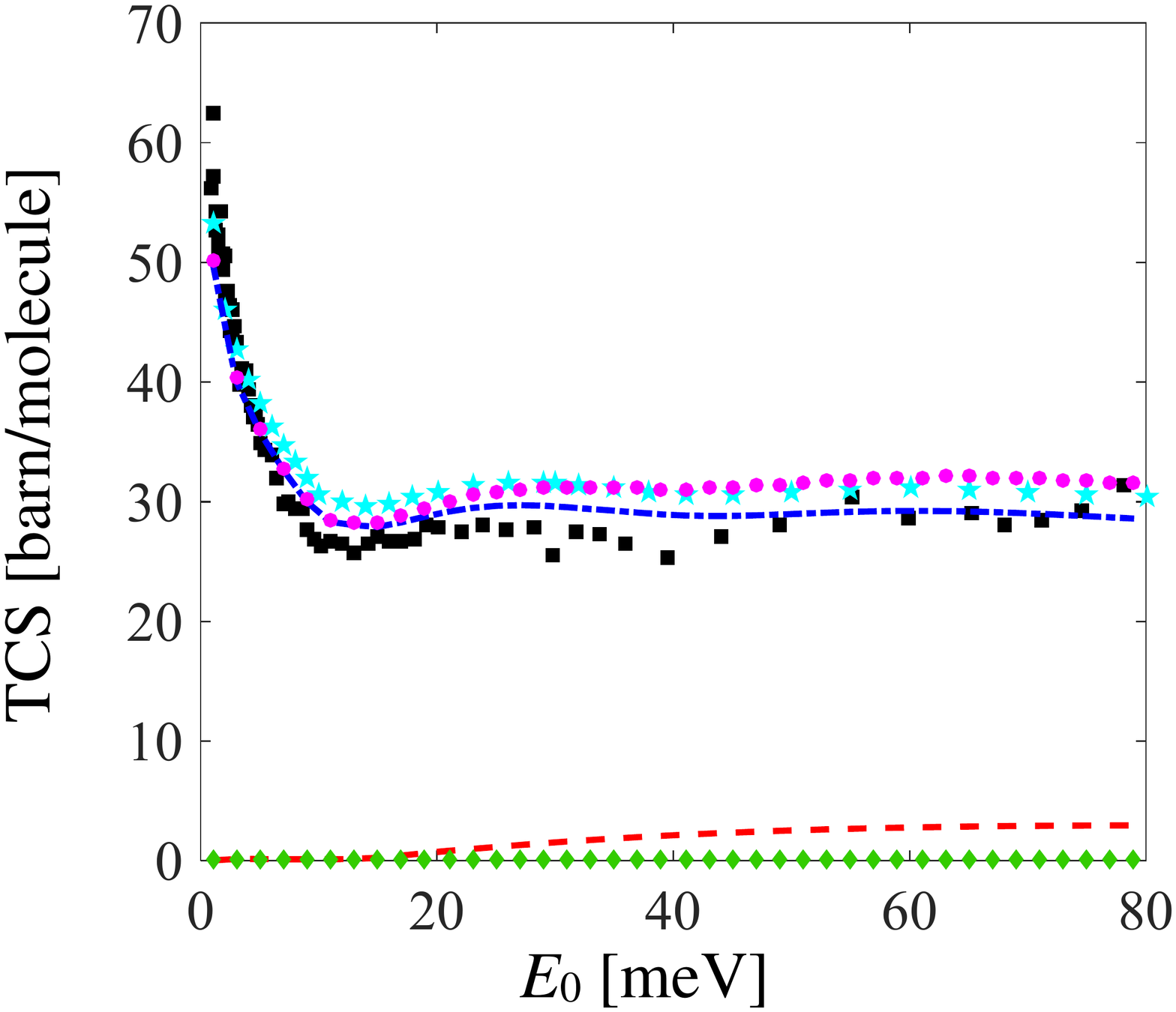}}
\caption{Energy dependence of the neutron total cross section measured by Seiffert \cite{Seiffert_rep} (black full circles). Calculations for the sample composition reported in Ref.\ \cite{Seiffert} are shown with pink full circles and compared with the result for pure HD (cyan stars) already displayed in Fig.\ \ref{TCS}. The individual contributions to the TCS are also separately shown: 94$\%$ HD (blue chain curve), 5.5$\%$ para-H$_2$ (red dashed curve) and 0.5$\%$ D$_2$ (green full diamonds).}
\label{mixture} 
\end{figure}

Unexpectedly, agreement between data and calculations is not as good as the one found for normal H$_2$ and D$_2$ \cite{Crisp_report}, not even at the high $E_0$ values where IG behaviour is foreseen to approximately hold. Therefore, in certain $E_0$ ranges, the observed discrepancies cannot be ascribed to the chosen modeling of $S_{\rm s}(Q,\omega)$ or to the neglect of a distinct dynamics. A nonnegligible problem in such comparisons is of course the unknown accuracy of the measured data, which unfortunately are provided without any estimate of the errors. Despite this serious lack of information, one can anyway try to explore the reasons of the mismatch in the HD case. For instance, the presence of a small fraction of ortho-H$_2$ was documented in Refs.\ \cite{Grammer2015, Guarini2015} to have been present in Seiffert's measurements on para-H$_2$. A possible effect of impurities might therefore be considered in the HD case as well. Information on this can be found in another paper by the author \cite{Seiffert} which reports the following composition determined by mass spectrometry: 94$\%$ HD, 5.5$\%$ H$_2$ and 0.5$\%$ D$_2$. However, normal H$_2$ cannot be assumed to be part of the mixture, since even a small amount would provide TCS values higher than those calculated for pure HD. Therefore we assumed the presence of the mentioned percentage of para-H$_2$. Figure \ref{mixture} shows that by considering the reported sample composition a small improvement is obtained only in the region around the first minimum of the TCS. More importantly, disagreement is even larger at high $E_0$, where free particle dynamics is reached and no doubt can concern the adequacy of the chosen IG model for HD. Therefore the sample composition is not the main source of the discrepancies.

\begin{figure}
\resizebox{0.9\textwidth}{!}
{\includegraphics[trim=0cm 0cm 2cm 0cm]{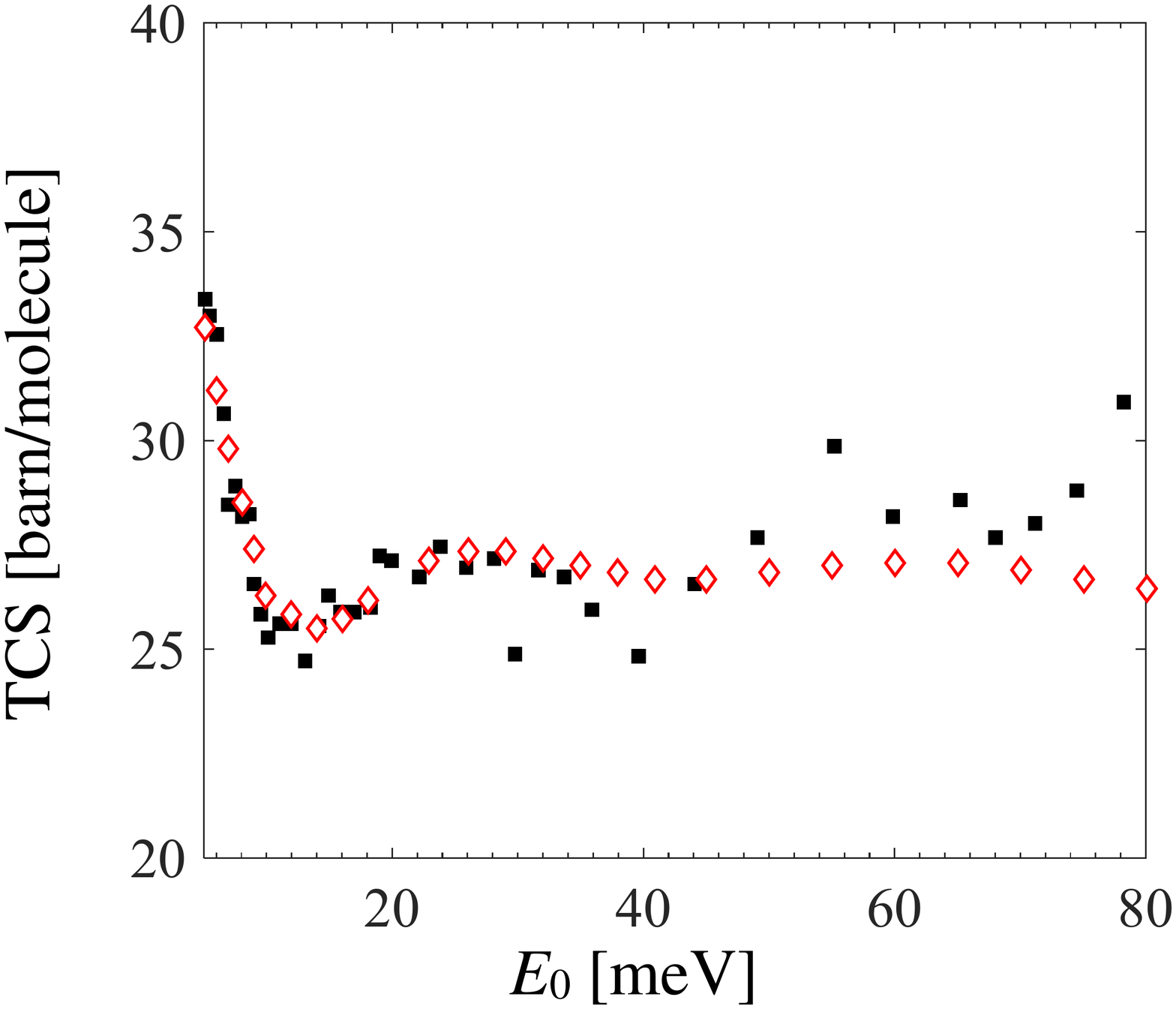}}
\caption{As in Fig.\ \ref{TCS} but considering a 15$\%$ lower value of the incoherent cross section of the H nucleus for the calculations (red empty diamonds).}
\label{lowerbH} 
\end{figure}

Another possible reason might be that also Seiffert's data witness the $\sim 30\%$ lower value of the H to D cross section ratio found from Compton neutron scattering on HD and on H$_2$-D$_2$ mixtures \cite{Chatzi2005}, and mainly attributed to an anomalous reduction of the H cross section like the one found from deep inelastic measurements on crystalline HCl \cite{Senesi}. In order to check this second (remote) possibility, we repeated our calculations for HD using diminished values of the (predominant) incoherent cross section of the H nucleus, finding that a $30 \%$ reduction is indeed too much, since it leads to a disagreement with experiment opposite in sign with respect to that of Fig.\ \ref{TCS}. Differently, a $\sim 15\%$ reduction of the H cross section (corresponding to a scattering length $b_{\rm inc,1}=23.3$ fm) seems to account fairly well for the measured TCS in and above the thermal region, as shown in Fig.\ \ref{lowerbH}. This finding is difficult to judge: on the one hand it might be a fortuitous result, with the H cross section playing only the role of an adjustable parameter that brings to a good coincidence data and calculations which otherwise differ for other (unclear) reasons that need further investigations. On the other hand, it might also be that H scatters in an anomalous way in the heteronuclear version of the molecule, but the lower reduction of its cross section found in the present case would then indicate that such an effect is strongly dependent on the incident energy. In truth, the first possibility seems more reasonable.

As a general remark, it is worth observing that the experimental data are affected by too marked oscillations and spurious scattering of the points above 30 meV, as it also happens for the D$_2$ measurements (see Fig.\ 5 of Ref.\ \cite{Guarini2016}). The accuracy of the data seems therefore somewhat dependent on the incident energy, and worsening as $E_0$ grows. On the other hand, at very low energies (e.g. below 10 meV), where deviations from IG behaviour have been shown to be important both in para-H$_2$ \cite{Guarini2015} and in D$_2$ \cite{Guarini2016}, one must recall the inadequacy of the oversimplified IG schematisation adopted for $S(Q,\omega)$, which for HD will be overcome only when quantum simulations of the VAF and determinations of $S(Q)$ become available also for this system. Therefore, at the present stage, the range of reliability for comparisons between data and calculations is actually rather limited (e.g. 10 $\le E_0 \le $ 30 meV). In the light of these considerations, the agreement of Fig.\ \ref{lowerbH} is therefore quite satisfactory, within the uncertainties about the measurements and the present limitations concerning the modeling of the translational motion of this molecule. 

\begin{figure}
\resizebox{1\textwidth}{!}
{\includegraphics[trim=-2cm 10cm 0cm 2cm]{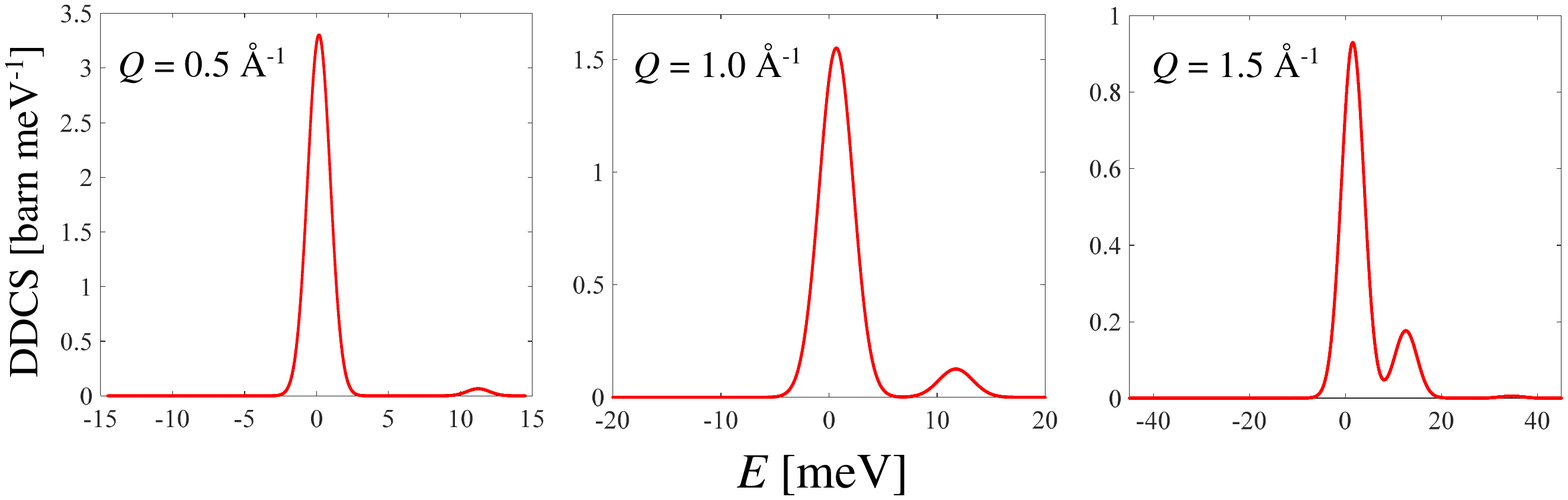}}
\caption{Double differential cross section of HD at an incident neutron energy of 80 meV and at three example $Q$ values.}
\label{spettri} 
\end{figure}

\begin{figure}
\resizebox{1\textwidth}{!}
{\includegraphics[trim=0cm 5cm 0cm 2cm]{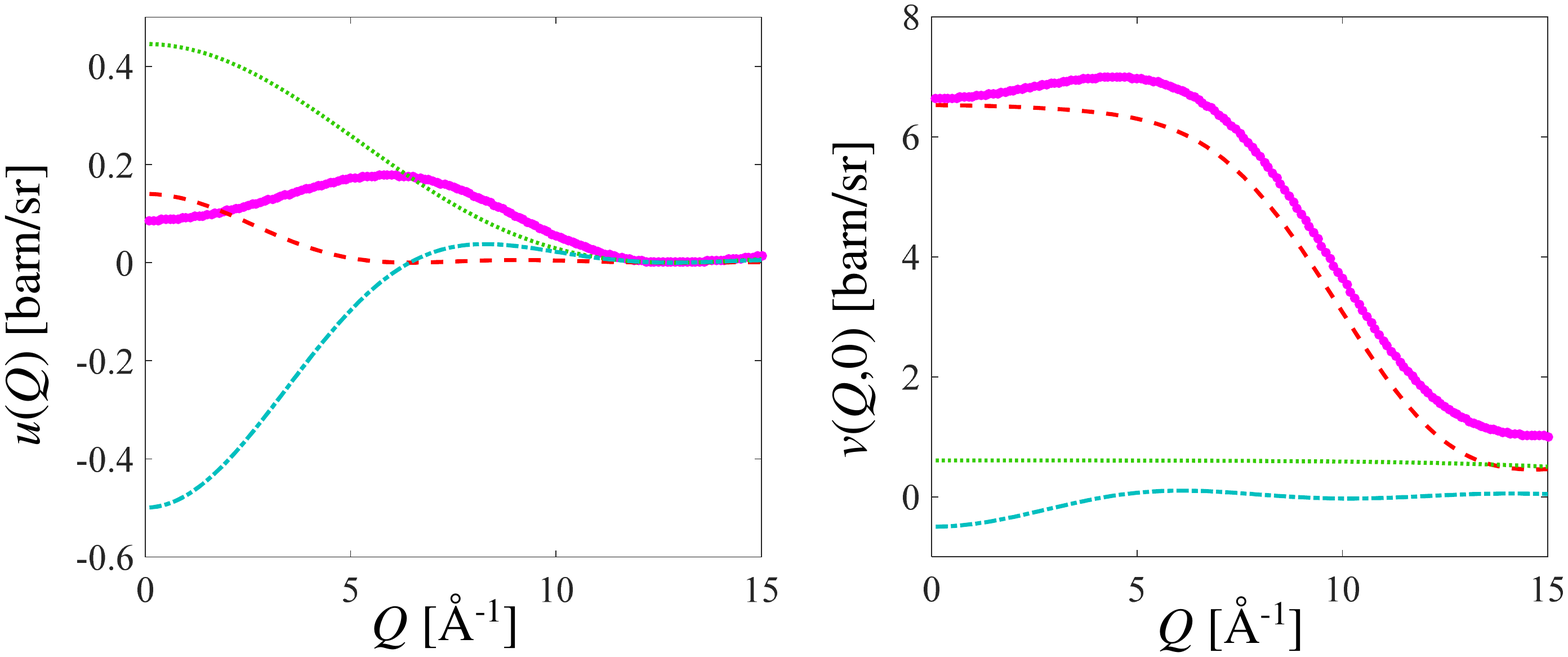}}
\caption{Left panel: intermolecular form factor $u(Q)$ (pink curve) according to Eq.\ (\ref{finalu}). Right panel: intramolecular form factor $v(Q,0)$ (pink curve) from Eq.\ (\ref{vQt2}) calculated at $t=0$. In both panels, the three different terms of the quoted equations are also plotted: the contribution due to H is the dashed red curve, the one due to D is the green dotted curve, and the cross HD term is the cyan chain curve; their sum provides the pink curves.}
\label{u&v} 
\end{figure}

Before drawing our overall conclusions, it can be interesting to have a snapshot of the typical DDCS spectra of HD at an incident energy (e.g. 80 meV) where an IG based calculation can be considered to be valid for a predominantly incoherent sample as HD, although with the mentioned uncertainty regarding the effective cross section of H. The following figures are all based on the nominal value of the H nuclear cross section. Figure \ref{spettri} gathers the HD spectra at some selected $Q$ values. Finally, for possible application in auspicable neutron diffraction and spectroscopic measurements on liquid HD we also show in Fig.\ \ref{u&v} the general features of the intra- and intermolecular form factors, $v(Q,0)$ and $u(Q)$, as a function of the exchanged wavevector $Q$. The various terms contributing, respectively, to Eq.\ (\ref{vQt2}) calculated at $t=0$ and to Eq.\ (\ref{finalu}) are also displayed separately.  
 
 \section{\label{concl} Conclusions} 

This work completes our review about the calculation of the neutron double differential cross section of diatomic molecules, providing formulas for a heteronuclear vibrating rotor. The interest for a formal treatment also of the heteronuclear case, and particularly for hydrogen deuteride, was triggered by possible applications of this fluid at low temperatures in neutron moderation. Unfortunately, the lack of experimental DDCS data for HD prevents one from a stringent test of the calculations. The present comparison with the (only available) total cross section data of liquid HD apparently suggests that the heteronuclear isotope of molecular hydrogen does not scatter as expected, although at the incident energies considered in this work the deviations, attributed to the cross section of the H nucleus, are not as large as those reported in the literature from very high energy experiments. Due to the unknown accuracy of the existing TCS experimental results, it is difficult to take the present unexpected findings as completely reliable, so it emerges very clearly that new accurate TCS measurements, as those done in recent years for para-H$_2$, are highly auspicable also for HD. It is also clear that a full assessment of our capability to predict the neutron double differential cross section of this criogenic liquid is subordinate to important inelastic scattering experiments and quantum simulation work aimed at a better modeling of the translational dynamics, as well as to neutron diffraction determinations of the static structure factor of HD. 

\section{Acknowledgments}

The author warmly thanks Ubaldo Bafile and Daniele Colognesi for useful advices and their critical reading of the manuscript.
This research was funded by Ministero dell'Istruzione dell'Universit\`a e della Ricerca Italiano (Grant No. PRIN2017-2017Z55KCW).

\newpage

\Table{\label{Tab1}Basic quantities used in the present calculations for HD} 
\br
Parameter&Description\\
\mr
$B$=5.538 meV \cite{Huber}&Rotational constant\\
$D$=0.003196  meV \cite{Huber}& Centrifugal distortion coefficient\\
$\hbar \omega_{\rm v}$=450.33 meV \cite{Huber} &Quantum of vibrational energy\\
$r_{\rm eq}$=0.74142 \AA~ \cite{Huber} & Equilibrium internuclear distance\\
$m_1$=1.00794 a.m.u. & Mass of the proton\\
$m_2$=2.01410 a.m.u & Mass of the deuteron\\
$\gamma_1=\frac{2}{3}$ & Fraction of the internuclear distance pertaining to the H nucleus\\
$\gamma_2=\frac{1}{3}$ & Fraction of the internuclear distance pertaining to the D nucleus\\
$b_{\rm coh,1}$=-3.7406 fm \cite{SearsCS}& Coherent scattering length of the H nucleus\\
$b_{\rm inc,1}$=25.274 fm \cite{SearsCS}& Incoherent scattering length of the H nucleus\\
$b_{\rm coh,2}$=6.674 fm \cite{SearsCS}& Coherent scattering length of the D nucleus\\
$b_{\rm inc,2}$=4.033 fm \cite{SearsCS}& Incoherent scattering length of the D nucleus\\
\br
\endTable

\end{document}